\documentclass[10pt]{article}
\def\@aabuffer{}
\def\ps@genrelativ{\let\@mkboth\@gobbletwo
\def\@oddhead{\hbox{}\sl\rightmark \hbox{} \sc{} \hfill \rm\thepage}
\def\@oddfoot{}
\def\@evenhead{\rm\thepage\sl\leftmark\hbox{} \hfill \sc{blablabla}\rm}
\def\@evenfoot{}
\def\sectionmark##1{}\def\subsectionmark##1{}}
\def\@aabuffer{}
\def\abstracts#1{
\begin{center}
{\begin{minipage}{4.2truein}
                 \footnotesize
                 \parindent=0pt #1\par
                 \end{minipage}}\end{center}
                 \vskip 2em \par}

\title{Background-Quantum Split Symmetry and \\Phase-Space Path-Integrals 
  \footnote{Talk given at the $ 6^{\rm th}$ International Conference on
    Path-Integrals from peV to TeV, Florence, Italy, 25-29 August 1998;
    to appear in the proceedings.}  \raisebox{4cm}[0mm][0mm]{
\begin{minipage}[b]{0cm} 
\normalsize \noindent \mbox{\hspace{2cm} hep-th/9812178}\\ \noindent
\mbox{\hspace{2cm} MZ-TH 98-51} 
\end{minipage}} }

\author{\normalsize M. Reuter\\ \normalsize \sl Institut f\"ur Physik,
  Universit\"at Mainz, D-55099 Mainz,
  Germany\\ \normalsize \sl E-mail: reuter@thep.physik.uni-mainz.de}

\date{ }

\begin{document}

\maketitle

\vspace{-1cm}
\abstracts{Phase-space path-integrals are used in order to illustrate
  various aspects of a recently proposed interpretation of quantum
  mechanics as a gauge theory of metaplectic spinor fields.}

\vspace{-1cm}
\section{\large Introduction}
 
In the framework of Hilbert bundles quantum mechanics can be
reformulated as a Yang-Mills theory over a symplectic manifold $ ({\cal
M},\omega) $, with an infinite dimensional gauge group and a
nondynamical connection \cite{metaqm}. The ``matter fields'' in this
gauge theory (metaplectic spinors) are local generalizations of states
and observables. They assume values in a family of local Hilbert spaces
(and their tensor products) which are attached to the points of the
phase-space ${\cal M}$. In this approach the rules of canonical
quantization are replaced by two new postulates with a very simple group
theoretical and differential geometrical interpretation. The first one
relates classical mechanics and semiclassical quantum mechanics while
the second one, invariance under the background-quantum split symmetry,
constructs the exact quantum theory by consistently sewing together an
infinity of local semiclassical expansions. In the following we
illustrate some features of this theory by means of hamiltonian
path-integrals. (For simplicity ${\cal M}$ is taken to be the symplectic
plane in these notes.)

\section{\large Semiclassical Quantum Mechanics from Representation Theory}

In our approach the passage from classical mechanics to semiclassical
quantum mechanics is based upon the following facts and constructions:

{\bf \noindent (1)} There exists a notion of spin bundles and spinor
fields appropriate for phase-spaces. Under local frame rotations these
so-called metaplectic spinors transform according to the spinor
representation of ${\rm Sp}(2{\mathit N})$, the analog of the Lorentz
group, or more precisely, of its double covering ${\rm Mp}(2{\mathit
N})$. Given a set of ``$\gamma$-matrices'' satisfying the symplectic
Clifford algebra $\gamma^{a}\gamma^{b}-\gamma^{b}\gamma^{a}=2\, i\,
\omega^{ab}$, the generators of ${\rm Mp}(2{\mathit N})$ are
$\Sigma^{ab}=\frac{1}{4} (\gamma^{a}\gamma^{b}+\gamma^{b}\gamma^{a})$.

{\bf \noindent (2)} Consider an auxiliary quantum system with $N$
canonically conjugate position and momentum operators $\widehat x^{i}$
and $\widehat \pi^{i}$ acting on the Hilbert space $\mathcal{V}$. The
canonical commutation relations imply that $\gamma^{a}\equiv \sqrt{2}
(\widehat \pi^{i}, \widehat x^{i})$ realizes the Clifford algebra and
gives rise to an infinite dimensional, unitary representation of ${\rm
Mp}(2{\mathit N}) $ on $\mathcal{V}$. The components of a spinor $\psi
\in \mathcal{V}$ in the $\widehat x$-eigenbasis are $\psi^{x}\equiv
\langle x| \psi\rangle$ with a ``spinor index'' $x\in
\mathbf{R}^{N}$. Metaplectic spinor fields $\psi^{x}(\phi^{a})$ are
fields on ${\cal M}$ which assume values in $\mathcal{V}$. $(
\phi^{a}\equiv (p^{i},q^{i}) $ are Darboux coordinates on ${\cal M}$.)
We write $\psi^{x}(\phi)\equiv \langle x| \psi\rangle_{\phi}$ in order
to indicate that $| \psi\rangle_{\phi}$ ``lives'' in the local Hilbert
space ${\cal V}_{\phi}$, a copy of $\cal V$ attached to the point
$\phi$.

{\bf \noindent (3)} Every hamiltonian vector field $h^{a}=\omega^{ab}\,
\partial_{b}H$ gives rise to a Lie-derivative
$\ell_{H}\psi=h^{a}\partial_{a}\psi+\frac{i}{2}\partial_{a}\partial_{b}H
\Sigma^{ab}\psi$ and similarly for arbitrary spinors/tensors
$\chi^{x...a...}_{y...b...}(\phi)$. Their Lie-transport along the
hamiltonian flow is governed by the equation $-\partial_{t}\chi
(\phi,t)=\ell_{H}\,\chi (\phi,t)$. In particular we consider singular
fields $\psi^{x}(\phi,t)=\eta^{x}(t)\,\delta(\phi-{\bf \Phi}_{\rm cl}
(t))$ localized along solutions ${\bf\Phi}_{\rm cl} (t)$ of Hamilton's
equation $\partial_{t}{\bf \Phi}^{a}_{\rm cl}=h^{a}({\bf \Phi}_{\rm
cl})$. The resulting evolution equation for the ``world line spinor''
$\eta$ reads $i \partial_{t} \eta=\frac{1}{2} \partial_{a}\partial_{b}
H({\bf \Phi}_{\rm cl} (t)) \Sigma^{ab} \eta$.  It has the same structure
as the classical Jacobi equation
$\partial_{t}\delta\phi^{a}=\partial_{b}h^{a}({\bf \Phi}_{\rm cl})\delta
\phi^{b}$ but with the vector representation of ${\rm Sp}(2{\mathit N})$
replaced by its spinor representation.

{\bf \noindent (4)} Given a hamiltonian $H$ we consider the standard
phase-space path-integral $\langle q_{2},t_{2}|
q_{1},t_{1}\rangle_{H}=\int \mathcal{D} p(t)\int \mathcal{D} q(t)\, {\rm
exp}\left[i\int_{t_{1}}^{t_{2}} dt \lbrace p^{i}\dot q^{i}-H(p,q)\rbrace
\right]$
\noindent over paths $\phi^{a}(t)\equiv (p^{i}(t),q^{i}(t))$ satisfying
$q(t_{1,2})=q_{1,2}$.  We shift $\phi(t)$ by an arbitrary classical
trajectory ${\bf \Phi}_{\rm cl}(t)\equiv(p_{\rm cl}(t),q_{\rm cl}(t))$,
i.e.  $\phi^{a}(t)={\bf \Phi}_{\rm cl}^{a}(t)+\varphi^{a}(t)$ with the
``fluctuation'' $\varphi(t)\equiv (\pi(t),x(t))$. Without any
approximation or assumption about the terminal points of $\bf{\Phi}_{\rm
cl}$, we obtain $ \langle q_{2},t_{2}| q_{1},t_{1}\rangle_{H}={\rm
exp}\, \left[i \left( S_{\rm cl}+p^{i}_{\rm cl}(t_{2})x_{2}^{i}-p_{\rm
cl}^{i}(t_{1})x_{1}^{i}\right)\right] K_{\mathcal{H}}$.  Here
$K_{\mathcal{H}}\equiv \langle x_{2},t_{2}|
x_{1},t_{1}\rangle_{\mathcal{H}}$ is the amplitude related to the
shifted integral $ K_{\mathcal{H}}=\int \mathcal{D}\pi\int \mathcal{D}
x\, {\rm exp}\left[ i \int_{t_{1}}^{t_{2}} dt \lbrace \pi^{i}\dot
x^{i}-\mathcal{H}(\varphi;{\bf\Phi}_{\rm cl})\rbrace \right] $
\noindent with boundary conditions $x(t_{1,2})=x_{1,2}\equiv
q_{1,2}-q_{\rm cl}(t_{1,2}).$ It contains the hamiltonian
$\mathcal{H}(\varphi;{\bf\Phi}_{\rm cl})\equiv H({\bf\Phi}_{\rm
cl}+\varphi)-\varphi^{a}\partial_{a} H({\bf\Phi}_{\rm cl})-
H({\bf\Phi}_{\rm cl})$. The Schr\"odinger equation equivalent to the
shifted path-integral reads $[i
\partial_{t}-\mathcal{H}(\widehat\varphi;{\bf\Phi}_{\rm cl})]\langle
x,t| x_{1},t_{1}\rangle_{\mathcal{H}}=0.$

{\bf \noindent (5)} We identify the operators $\widehat
\varphi^{a}\equiv (\widehat \pi^{i},\widehat x^{i})$ appearing in the
canonical quantization of the fluctuations $\varphi^{a}(t)$ with those
of the auxiliary system which were introduced in order to represent the
Clifford algebra.

{\bf \noindent (6)} In the semiclassical approximation one has
  $\mathcal{H}(\widehat \varphi, {\bf\Phi}_{\rm
  cl})=\frac{1}{2}\partial_{a}\partial_{b}H({\bf\Phi}_{\rm
  cl})\widehat\varphi^{a}\widehat\varphi^{b}+ {\rm O}(\widehat
  \varphi^{3})$ which is in the Lie algebra of ${\rm Mp}(2{\mathit N})$.
  This implies that $\eta^{x}(t)\equiv\langle x,t|
  x_{1},t_{1}\rangle_{\mathcal{H} }$ is a world-line spinor: the
  equation of motion given in (3) coincides exactly with the
  semiclassical approximation of the Schr\"odinger equation in (4).

\noindent The upshot of the preceding arguments is that the notion of
semiclassical wave functions follows from the classical concept of the
Jacobi fields by simply passing over from the vector to the spinor
representation of ${\rm Sp}(2{\mathit N})$.

\section{\large Exact Quantum Mechanics from Split Symmetry}

\noindent Within this Hilbert bundle approach, the natural description
of semiclassical quantum mechanics is in terms of singular spinor fields
with support along ${\bf\Phi}_{\rm cl}(t)$ only. The exact theory is
formulated in terms of smooth fields $\psi^{x}(\phi)=\langle x|
\psi\rangle_{\phi}$ defined everywhere on ${\cal M}$. Loosely speaking,
a single field $\psi^{x}(\phi)$ encapsulates the information contents of
the world line spinors along the totality of {\it all} classical paths.
Heuristically, this can be understood as follows. Let us assume we know
the solutions $\eta(t)$ of the semiclassical Schr\"odinger equation for
a congruence of neighboring classical trajectories $\Phi(t)$. We expect
the wave function $\eta_{1}$ belonging to some trajectory $\Phi_{1}$ to
provide a reasonable approximation to the complete theory within a
tubular neighborhood of $\Phi_{1}(t)$. Furthermore, let us suppose that
there is a nearby solution $\Phi_{2}(t)$ such that the tubular
neighborhood within which its wave function $\eta_{2}(t)$ is valid
overlaps with the one of $\Phi_{1}$. As a consequence, there is a region
in phase-space where both semiclassical expansions apply and where their
predictions must agree. For instance, the expectation value of the
``position'' is given by $\Phi_{1}^{a}(t)+\bar
\eta_{1}(t)\,\widehat\varphi^{a}\,\eta_{1}(t)$ according to the first,
and by $\Phi^{a}_{2}(t)+\bar
\eta_{2}(t)\,\widehat\varphi^{a}\,\eta_{2}(t)$ according to the second
expansion. (Here $\bar \eta\,\widehat\varphi^{a}\,\eta\equiv \int dx\int
dy\,\,\eta^{\ast}_{x}(\widehat \varphi^{a})^{x}_{\,\, y}\,\eta^{y}$.)
Invariance under the background-quantum split symmetry means that these
two values coincide.

\noindent The time-independent spinor field $\psi(\phi)=|
\psi\rangle_{\phi}$ which describes this situation in the exact theory
is constructed as follows. Since at different times $t$ the world-line
spinor $\eta(t)\in {\cal V}_{\Phi(t)}$ lives in different local Hilbert
spaces along the trajectory $\Phi(t)$ we define the value of
$\psi(\phi)$ along $\phi=\Phi(t)$ by setting
$|\psi\rangle_{\Phi(t)}=\eta(t)$ for all $t$. Then, in order to find
$\psi(\phi)$ in the directions transverse to $\Phi(t)$, we repeat this
construction for all pairs $(\Phi,\eta)$ which are connected by the
requirement of the split symmetry. It can be shown \cite{metaqm} that
this procedure leads to a spinor field $\psi^{x}(\phi)$ which satisfies
an equation of the form $(\partial_{a}+i\widetilde \Gamma_{a})\psi=0$
where $\widetilde\Gamma_{a}$ is a set of hermitian operators on
$\mathcal{V}$. In the gauge theory formulation \cite{metaqm} this
condition means that $\psi$ is covariantly constant with respect to a
universal connection (gauge field) $\widetilde\Gamma_{a}$. This
connection can be found by looking at how the shifted path-integral
responds to the replacement ${\bf\Phi}_{\rm cl}\rightarrow
{\bf\Phi}_{\rm cl} +\delta\phi$ where $\delta\phi\equiv(\delta p_{\rm
cl},\delta q_{\rm cl})$ is a Jacobi field. The result reads $\delta
K_{\mathcal{H}}=[\delta q_{\rm cl}(t_{2})\partial_{x_{2}}+\delta q_{\rm
cl}(t_{1})\partial_{x_{1}}-i x_{2} \delta p_{\rm cl}(t_{2})+i x_{1}
\delta p_{\rm cl}(t_{1})]K_{\mathcal{H}}$.
\noindent This equation describes an infinitesimal parallel transport of
the spinor $K_{\mathcal{H}}$ with the connection
$\widetilde\Gamma_{a}$. Here $K_{\mathcal{H}}$ is considered a function
of $x_{2},t_{2}$ with $x_{1},t_{1}$ fixed. Thus we read off that the
universal connection $\widetilde \Gamma$ is given by $\widetilde
\Gamma_{a}=\omega_{ab}\,\widehat\varphi^{b}$ which happens to be
independent of $\phi$ for a flat phase-space.

\noindent We interpret $\langle \widehat
\phi^{a}\rangle=\phi^{a}+_{\phi}\langle
\psi|\widehat\varphi^{a}|\psi\rangle_{\phi}$ as the exact quantum
mechanical expectation value of the position in phase-space. The
requirement of the split symmetry means that this value is the same for
all pairs $(\phi, |\psi\rangle_{\phi})$. When we go from the point
$\phi$ to another point $\phi+\delta \phi$ the state $|
\psi\rangle_{\phi}\in {\cal V}_{\phi}$ is replaced by a new state $|
\psi\rangle_{\phi+\delta\phi}\in {\cal V}_{\phi+\delta\phi}$ in such a
way that the change of the $\widehat \varphi$-expectation value
precisely cancels the change of the classical contribution,
$\delta\phi^{a}$. This condition is fulfilled if the state vector
changes by an amount $\delta | \psi\rangle_{\phi}=-i\,
\delta\phi^{a}\widetilde\Gamma_{a}|\psi\rangle_{\phi}$, i.e. if it is
parallel-transported with the connection $\widetilde \Gamma$. This
parallel transport is ``almost integrable'' . Since the curvature of
$\widetilde \Gamma_{a}$ is $\omega_{ab}$ times the unit operator on
$\cal V$, only the (irrelevant) phase of the transported spinor is
path-dependent.
     
\noindent Classical observables $f(\phi)$ are represented by covariantly
constant operators $\mathcal{O}_{f}(\phi)$ which generalize
$\phi^{a}+\widehat \varphi^{a}$ above. For $\cal M$ flat one has
$\mathcal{O}_{f}(\phi)=f(\phi+\widehat\varphi)$. At the level of the
$f$'s the operator product of the $\mathcal{O}$'s corresponds to the
star product of the Weyl symbol calculus.

\noindent The equivalence of the gauge theory approach with the standard
formulation of quantum mechanics is established by noting that the
parallel transport with $\widetilde \Gamma$ can be used to identify all
local Hilbert spaces ${\cal V}_{\phi}$ with a single reference Hilbert
space at $\phi=\phi_{0}$, say. This reference Hilbert space is the one
used in the standard approach and $\psi^{x}(\phi_{0})\equiv\Psi(x)$,
regarded as a function of $x$, is the ordinary wave function.

\noindent Time evolution is described by a local unitary frame rotation:
$i \partial_{t}\psi(\phi)=\mathcal{O}_{H}(\phi)\psi(\phi)$.  This
fundamental dynamical law unifies two rather different types of time
evolution: the Lie-transport along the hamiltonian flow and the standard
Schr\"odinger equation.

\noindent For further details and the generalization of arbitrary curved
phase-spaces we refer to the literature \cite{metaqm}. One of the
central results is that the structures of quantum mechanics emerge from
those of classical mechanics by (i) switching from the vector to the
spinor representation of ${\rm Sp}(2{\mathit N})$ and (ii) imposing the
split symmetry. On curved phase-spaces the latter involves a
quantum-deformed symplectic analog of the exponential map.

\   \\
\   \\

\end{document}